\newcommand{\fig}[1]{\mbox{Figure\hspace{0.2em}\ref{#1}}}
\newcommand{\eqn}[1]{\mbox{Equation\hspace{0.2em}\ref{#1}}}
\newcommand{\sect}[1]{\mbox{\S\ref{#1}}}
\newcommand{\tbl}[1]{\mbox{Table\hspace{0.3em}\ref{#1}}}
\newcommand{\aeff}{\Omega_\mathrm{eff}}

\newcommand{\np}{N_\mathrm{P}}
\newcommand{\nt}{N_\mathrm{T}}
\newcommand{\zt}{z_\mathrm{t}}

\documentclass[iop]{emulateapj-rtx4}

\usepackage{graphicx}
\usepackage{natbib}
\usepackage{amssymb}

\begin{document}
\title{The TAOS Project: Results From Seven Years of Survey Data}
\journalinfo{Submitted to Astronomical Journal}
\submitted{Submitted to Astronomical Journal 2013 January 16}

\shorttitle{TAOS Seven Year Results}
\shortauthors{Zhang et al.}

\author{
Z.-W.~Zhang\altaffilmark{1},
M.~J.~Lehner\altaffilmark{1,2,3},
J.-H.~Wang\altaffilmark{1},
C.-Y.~Wen\altaffilmark{1},
S.-Y.~Wang\altaffilmark{1},
S.-K.~King\altaffilmark{1},
\'{A}.~P.~Granados\altaffilmark{4},
C.~Alcock\altaffilmark{3},
T.~Axelrod\altaffilmark{5},
F.~B.~Bianco\altaffilmark{6},
Y.-I.~Byun\altaffilmark{7},
W.~P.~Chen\altaffilmark{8},
N.~K.~Coehlo\altaffilmark{9},
K.~H.~Cook\altaffilmark{1},
I.~de~Pater\altaffilmark{10},
D.-W.~Kim\altaffilmark{11},
T.~Lee\altaffilmark{1},
J.~J.~Lissauer\altaffilmark{12},
S.~L.~Marshall\altaffilmark{13},
P.~Protopapas\altaffilmark{14,3},
J.~A.~Rice\altaffilmark{9}, and
M.~E.~Schwamb\altaffilmark{15,16}
}
\altaffiltext{1}{Institute of Astronomy and Astrophysics, Academia Sinica.
11F of Astronomy-Mathematics Building, National Taiwan University.
No.1, Sec. 4, Roosevelt Rd, Taipei 10617, Taiwan}
\email{zwzhang@asiaa.sinica.edu.tw}
\altaffiltext{2}{Department of Physics and Astronomy, University of
 Pennsylvania, 209 South 33rd Street, Philadelphia, PA 19104}
\altaffiltext{3}{Harvard-Smithsonian Center for Astrophysics, 60 Garden Street,
 Cambridge, MA 02138}
\altaffiltext{4}{Instituto de Astronom\'{i}a, Universidad Nacional Aut\'{o}noma
  de M\'{e}xico, Apdo. Postal 106, Ensenada, Baja California, 22800 M\'{e}xico}
\altaffiltext{5}{Steward Observatory, 933 North Cherry Avenue, Room N204
 Tucson AZ 85721}
\altaffiltext{6}{New York University, Center for Cosmology and Particle
Physics, 4 Washington Place, New York, NY 10003.}
\altaffiltext{7}{Department of Astronomy and University Observatory,
  Yonsei University, 134 Shinchon, Seoul 120-749, Korea}
\altaffiltext{8}{Institute of Astronomy, National Central University, No. 300,
 Jhongda Rd, Jhongli City, Taoyuan County 320, Taiwan}
\altaffiltext{9}{Department of Statistics, University of California Berkeley,
 367 Evans Hall, Berkeley, CA 94720}
\altaffiltext{10}{Department of Astronomy, University of California Berkeley,
 601 Campbell Hall, Berkeley CA 94720}
\altaffiltext{11}{Max Planck Institute for Astronomy, K\"{o}nigstuhl 17, 
D-69117~Heidelberg, Germany}
\altaffiltext{12}{Space Science and Astrobiology Division 245-3, NASA Ames
 Research Center, Moffett Field, CA, 94035}
\altaffiltext{13}{Kavli Institute for Particle Astrophysics and Cosmology,
 2575 Sand Hill Road, MS 29, Menlo Park, CA 94025}
\altaffiltext{14}{Initiative in Innovative Computing, Harvard University,
 60 Oxford St, Cambridge, MA 02138}
\altaffiltext{15}{Department of Physics, Yale University, New Haven, CT 06511}
\altaffiltext{16}{Yale Center for Astronomy and Astrophysics, Yale University,
 P.O. Box 208121, New Haven, CT 06520}

\begin{abstract}
The Taiwanese-American Occultation Survey (TAOS) aims to detect
serendipitous occultations of stars by small ($\sim$1~km diameter)
objects in the Kuiper Belt and beyond. Such events are very rare
($<10^{-3}$ events per star per year) and short in duration
($\sim$200~ms), so many stars must be monitored at a high readout
cadence. TAOS monitors typically $\sim$500 stars simultaneously at a
5~Hz readout cadence with four telescopes located at Lulin Observatory
in central Taiwan. In this paper, we report the results of the search
for small Kuiper Belt Objects (KBOs) in seven years of data. No
occultation events were found, resulting in a 95\%~c.l. upper limit on
the slope of the faint end of the KBO size distribution of $q = 3.34$
to 3.82, depending on the surface density at the break in the size
distribution at a diameter of about 90~km.
\end{abstract}


\section{Introduction}
\label{sec:intro}
The size distribution of Kuiper Belt Objects has been accurately
measured down to diameters of $D > 30$~km \citep{2009ApJ...696...91F,
  2008AJ....136...83F, 2008Icar..198..452F, 2008Icar..195..827F,
  2004AJ....128.1364B, 2002ARA&A..40...63L}, while
\citet{2009AJ....137...72F} have provided measurements down to
diameters $D \gtrsim 15$~km. The size distribution of large KBOs
approximately follows a power law of the form
\begin{equation}
\frac{dn}{dD} \propto D^{-4.5}
\end{equation}
down to a diameter of $D\approx 90$~km. A clear break in the size
distribution has been detected near 90~km \citep{2009ApJ...696...91F,
  2009AJ....137...72F, 2004AJ....128.1364B}, giving way to a shallower
distribution for smaller objects. Various models of the formation of
the Kuiper Belt have been developed \citep{2009P&SS...57..201B,
  2009ApJ...690L.140K, 2005Icar..173..342P, 2004AJ....128.1916K,
  2001AJ....121..538K, 1999Icar..142....5B, 1999AJ....118.1101K,
  1999ApJ...526..465K, 1997Icar..125...50D, 1996AJ....112.1203S,
  1995AJ....110.3073D} which predict this break in the size
distribution, but these models make vastly different predictions on
the size distribution for objects smaller than the break
diameter. These models are based on different scenarios of the
dynamical evolution of the Solar System and the internal structure of
the KBOs themselves. It is generally assumed that the outward
migration of Neptune stirred up the orbits of objects in the Kuiper
Belt, which subsequently underwent a process of collisional erosion,
giving rise to a shallower size distribution. The simplest of these
models also predict a small-end power law size distribution of the
form
\begin{equation}
\frac{dn}{dD} \propto D^{-q},
\end{equation}
where the slope $q$ depends primarily on the internal strength of the
KBOs. Solid objects held together by material strength give rise to
larger slopes, while the weaker gravitationally bound rubble piles are
more easily broken up and give rise to smaller values of $q$.  A
measurement of the size distribution of smaller objects (down to $D
\gtrsim 100$~m) is needed in order to constrain these models.
Furthermore, such a measurement would provide important information on
the origin of short-period comets \citep{2008ApJ...687..714V,
  2006Icar..182..527T, 1997Sci...276.1670D, 1997Icar..127...13L,
  1997Icar..127....1M, 1993AJ....105.1987H}.

The direct detection of such objects is difficult because they are
extremely faint, with typical magnitudes $R > 28$, and are thus
invisible to surveys using even the largest telescopes. However, a
small KBO will induce a detectable drop in the brightness of a distant
star when it passes across the line of sight to the star
\citep{2012MNRAS.tmp..367C, 2012ApJ...761..150S, 2010AJ....139.2003W,
  2010AJ....139.1499B, 2009Natur.462..895S, 2009AJ....137.4270B,
  2009AJ....138.1893W, 2009AJ....138..568B, 2008MNRAS.388L..44L,
  2008ApJ...685L.157Z, 2008AJ....135.1039B, 2007MNRAS.378.1287C,
  2007AJ....134.1596N, 2006Natur.442..660C, 2006AJ....132..819R,
  2003ApJ...594L..63R, 2003ApJ...589L..97C, 2003ApJ...587L.125C,
  2000Icar..147..530R, 1997MNRAS.289..783B, 1987AJ.....93.1549R,
  1976Natur.259..290B}. High speed photometry is needed to detect
these events, given that we expect a typical event duration of about
$200$~ms. Detection of such events is further complicated by the fact
that the sizes of the objects of interest are on the order of the
Fresnel scale (about 1.4~km for our median observed wavelength of
about 600~nm and a typical object distance of 43~AU), and the
resulting lightcurves exhibit significant diffration features
\citep{2007AJ....134.1596N}. The goal of the TAOS project is to detect
such occultation events and measure the size distribution of Kuiper
Belt Objects (KBOs) with diameters $0.5~\mathrm{km} < D <
30~\mathrm{km}$.

To date, 18~events consistent with occultations by objects in the
outer Solar System have been detected. The three detections claimed by
\citet{2006AJ....132..819R}, if they are occultations, are unlikely to
have been caused by KBOs (the distances are inconsistent). On the
other hand, two detections reported by \citet{2009Natur.462..895S} and
\citet{2012ApJ...761..150S} are consistent with occultation events by
1~km diameter KBOs at 45~AU. \citet{2012ApJ...761..150S} found one
additional event consistent with a KBO occultation, but they reject
this event because of its high inclination (it was detected at an
ecliptic latitude of $b = 81.5^\circ$). The remaining 12~possible
events \citep{2007MNRAS.378.1287C} were found in archival RXTE
measurements of Sco~X-1. However, the interpretation of these
detections as occultation events has been challenged
\citep{2008ApJ...677.1241J, 2009ApJ...701.1742B}.

\begin{deluxetable*}{lrrr}
\tablecolumns{4}
\tablewidth{0pc}
\tablecaption{Dataset parameters.}
\tablehead{ & \citet{2008ApJ...685L.157Z}  & \citet{2010AJ....139.1499B}
 & This Work}
\startdata
Start Date & 2005 Feb 7 & 2005 Feb 7 & 2005 Feb 7 \\ 
End Date & 2006 Dec 31 & 2008 Aug 2 & 2011 Sep 8 \\
Data Runs & 156 & 414 & 858 \\
Light-curve Sets\tablenotemark{a}
 & 110,554 & 366,083 & 835,732 (209,130) \\
Total Exposure (star--hours)\tablenotemark{a} &
152,787 & 500,339 & 1,159,651 (292,514) \\
Photometric Measurements\tablenotemark{a}
 & $7.8\times 10^9$ & $ 2.7\times 10^{10}$ &
 $6.7\times 10^{10}$ ($2.1\times10^{10}$) \\
Flux Tuples\tablenotemark{a} &
$2.6\times 10^9$ & $9.0\times 10^9$ & $2.1\times 10^{10}$ ($5.3\times 10^9$) \\
\enddata
\tablenotetext{a} {Values for four-telescope data shown in parentheses.}
\label{tbl:dataset}
\end{deluxetable*}

TAOS has been in operation since February 2005. The system is
described in detail in \citet{2009PASP..121..138L}. In summary, the
survey operates four 50~cm telescopes at Lulin Observatory in central
Taiwan. The telescopes are very fast (F/1.9), with a field of view of
3~deg$^2$. Each telescope is equipped with a camera which utilizes a
single 2k$\times$2k CCD imager. All four telescopes monitor the same
stars simultaneously at a readout cadence of 5~Hz.  We require that
any events be detected simultaneously in each of the telescopes in
order to minimize the false positive rate \citep{2010PASP..122..959L}.

High speed cadence is achieved using a special CCD readout mode called
\emph{zipper mode}, which is described in detail in
\citet{2009PASP..121..138L}. To summarize, instead of taking repeated
exposures, we leave the shutter open for the duration of a data
run. At a cadence of 5~Hz, we read out a subimage comprising 76~rows,
which we have termed a \emph{rowblock}. When the rowblock is read out,
the remaining photoelectrons on the CCD are shifted down by 76~rows as
well. If one considers starting with a rowblock at the far edge of the
CCD, photoelectrons are collected and then shifted to the next
rowblock. Photoelectrons are then collected in addition to those from
the first rowblock, and then they are shifted to the next. By the time
the original photoelectons are read out, photoelectrons have been
collected from every star on the focal plane in a single rowblock,
although at different epochs.

\begin{figure*}
\plotone{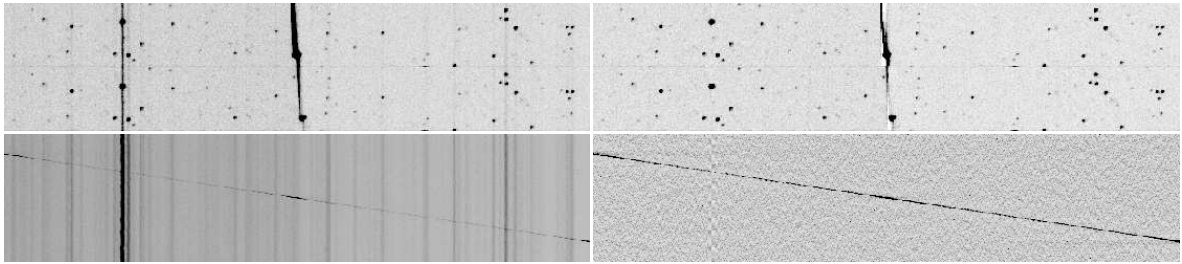}
\caption[]{Top left: Horizontal subsections of two consecutive
  rowblocks with a bright moving object in the center. (The boundary
  between the rowblocks is visible as a dark horizontal line bisecting
  the image. This is a slight excess in dark current from the edge of
  the CCD.) The vertical streaks (see text) from the brighter stars
  are also evident. Top right: The same rowblock subsections after the
  background streak subtraction. While the streaks from the bright
  stars are removed, the moving object itself is not completely
  removed, and the background flux near the moving object is
  over-subtracted. Bottom left: grayscale image representation of the
  background flux subtracted from the rowblocks near those shown in
  the top panels. Here each row represents the flux subtracted from a
  single rowblock, while the columns correspond to those in the top
  panels. The vertical streaks in this image are the streaks left by
  the stars during the zipper mode readout. The moving object in
  original data cause the background to be over-subtracted, and the
  over-subtracted flux clearly shows the trajectory of the moving
  object across the focal plane. Bottom right: The same background
  image from the bottom left panel after applying the high pass mean
  and variance filters to remove the streaks from the bright
  stars. The over-subtracted background flux from the moving object is
  clearly evident.}
\label{fig:obj}
\end{figure*}

While zipper mode readout allows us to image every star in the focal
plane, there are two characteristics of the data that reduce the
signal-to-noise ratio (SNR). The first is that sky background is
collected as a rowblock is shifted across the focal plane, while
photoelectrons from a single star are only collected at one
location. Furthermore, it takes about 1.3~ms to read out a single row,
corresponding to about 95~ms for an entire rowblock. So the actual
exposure of a star is only 105~ms, while sky background is collected
for a total of 5.4~s. Second, flux is also collected from the stars
during the readout stage, and the brighter stars will leave
significant numbers of photoelectrons in the pixels as they are being
shifted. This flux will appear as bright vertical streaks in the
images (see the upper left panel of \fig{fig:obj}). Nevertheless,
zipper mode is successful in collecting high SNR photometric data on
enough stars for the project to successfully probe a significant
fraction of the allowed size distribution parameter space.

In this paper we report on the analysis of seven years of zipper mode
data collected by the TAOS system. In \sect{sec:dataset}, we describe
the data set used in this analysis. In \sect{sec:pipeline}, we provide
a detailed discussion of the analysis pipeline. In \sect{sec:results},
we present the results of the analysis of the data set, and finally in
\sect{sec:plans}, we discuss our future plans for the survey.

Throughout the remainder of this paper, we use the following
definitions. A \emph{data run} is a series of high cadence
multi-telescope measurements on a single field, typically for
1.5~hours at a time. A \emph{lightcurve} is a time series of
photometric measurements of a single star from one telescope in one
data run, and a \emph{lightcurve set} is a set of multi-telescope
photometric measurements of a single star during one data run. A
\emph{flux tuple} is a set of multi-telescope measurements (with a
minimum of three telescopes) of one star at a single epoch. A
lightcurve set is thus a time series of flux tuples.

\section{The Seven Year Dataset}
\label{sec:dataset}

Earlier results from the TAOS search for small KBOs were reported in
\citet{2008ApJ...685L.157Z} and \citet{2010AJ....139.1499B}. These two
papers describe results obtained from operating three telescopes. The
dataset used in \citet{2010AJ....139.1499B} ended on 2008 August 2,
when the fourth telescope became operational. The addition of the
fourth telescope to the array was beneficial in two ways. First, using
four telescopes increases the significance of any candidate
occultation events, while the false positive rate remains constant
\citep{2010PASP..122..959L}. Second, we have experienced frequent
problems with the reliability of our cameras and telescopes. With four
telescopes, we can still operate with three telescopes when one of the
systems is shut down for repair. Given that a minimum of three
telescopes is necessary to keep the false positive rate low enough
\citep{2010PASP..122..959L}, we can still collect useful data in the
event that there is a problem with one of the systems. The overall
operational efficiency of the survey thus improved significantly since
we could still collect data when one telescope was offline.

The current dataset is summarized in \tbl{tbl:dataset}, along with the
datasets used in \citet{2008ApJ...685L.157Z} and
\citet{2010AJ....139.1499B}. The current dataset is more than
2.3~times larger than that used in \citet{2010AJ....139.1499B}. 90\%
of the photometric data in this set was acquired within 6$^\circ$ of
the ecliptic in order to maximize the event rate. The results quoted
in this paper are thus applicable to both the hot (inclination $i >
5^\circ$) and cold ($i < 5^\circ$) KBO populations.

\section{Data Analysis}
\label{sec:pipeline}

The data analysis took place in three stages: photometric reduction of
raw image data, the search for occultation events, and the detection
efficiency simulation. These stages roughly follow the same procedures
outlined in \citet{2010AJ....139.1499B}, but several improvements were
made to the pipeline in this round of data analysis. The new pipeline
is described in the following subsections.

\setcounter{footnote}{0}

\begin{figure*}
\plotone{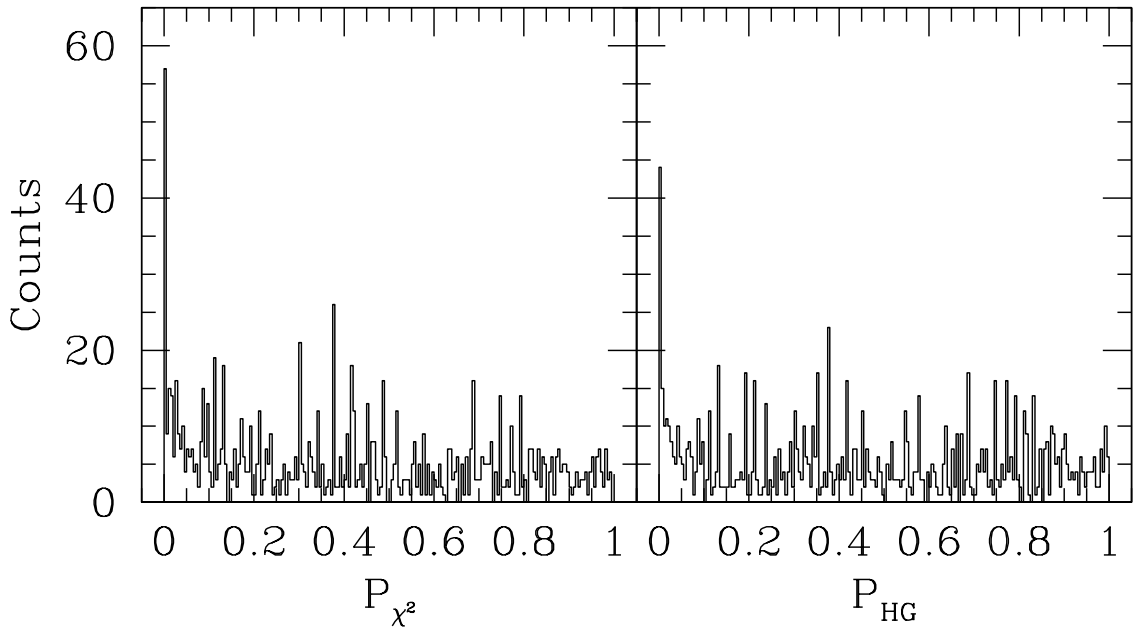}
\caption[]{Histograms of p-values for the Pearson's $\chi^2$ test
  (left panel) and hypergeometric test (right panel) for each of the
  data runs in the data set.}
\label{fig:stat}
\end{figure*}

\subsection{Photometric Reduction}
\label{sec:phot}
For the first step in the pipeline, a custom photometric reduction
pipeline \citep{2009PASP..121.1429Z} is used to measure the brightness
of each star in the field at each epoch, and the resulting
measurements are assembled into a time series lightcurve for each
star. For every star, the lightcurve from each telescope is combined
into a lightcurve set. Next, bad data points (e.g. measurements where
the star lies near the edge of the focal plane) are removed from the
lightcurves \citep[see][for details]{2009PASP..121.1429Z}.

A significant improvement added to this stage in the process is the
automated flagging of photometric measurements affected by bright
objects (such as satellites or airplanes) moving across the field of
view. As discussed in \citet{2010AJ....139.1499B}, several candidate
events were found in the previous dataset. Visual inspection of the
relevant images revealed that these events were caused by brigh moving
objects passing near the star that appeared to be occulted. The new
algorithm to detect these bright moving objects is described in
\sect{sec:movobj}.

\subsubsection{Bright Moving Object Detection}
\label{sec:movobj}
As mentioned in \sect{sec:phot}, bright objects moving across the
field of view during zipper mode image collection can give rise to a
large number of false positive events. This is caused by the
background subtraction in the photometric reduction. As discussed in
\sect{sec:intro}, the zipper mode readout has the disadvantage that
the brighter stars leave streaks along the columns in the images due
to the finite time it takes to read out a row from the CCD. These
streaks are removed during the photometric reduction by subtracting an
estimate of the mode of the column from each pixel in the column,
where the mode is calculated for each zipper mode rowblock
\citep{2009PASP..121.1429Z}. A bright moving object passing over a
particular column causes the mode of that column to be over-estimated,
and the background is thus over-subtracted, causing an artificial drop
in the measured brightness of any star whose aperture includes that
column. (see the top panels of \fig{fig:obj}).

In order to detect these events, we save the 3$\sigma$-clipped column
mean for each rowblock and column on all four cameras. These
background data are stored in FITS format, and an example is shown in
the bottom left panel of \fig{fig:obj}. Note that each column in this
image corresponds to an actual column on the CCD, but the row
corresponds to a rowblock in the zipper mode data. The pixel value is
simply the column mean (after 3$\sigma$ clipping) for that rowblock
and column. The streaks from the bright stars are evident in the
image. Also clearly visible is the object shown in the top panels
moving across the image. We note that what appears as a moving object
is actually the high value for the column average due to the moving
object. Nevertheless, it is straightforward to identify which columns
in which rowblocks are affected by the over-estimated background.

First, we normalize the image by applying the same rolling clipped
mean and variance filter to each of the columns we use to remove the
slowly varying trends along the lightcurves
\citep{2010PASP..122..959L}. That is, we replace each column average
value $b_{mn}$, where $m$ is the row block index and $n$ is the
column, with a normalized value $d_{mn}$, defined as
\begin{eqnarray}
c_{mn} & = & b_{mn} - \mu_{mn},\nonumber\\
d_{mn} & = & \frac{c_{mn}}{\sigma_{mn}},
\label{eq:roll}
\end{eqnarray}
where $\mu_{mn}$ is the local ($3\sigma$-clipped) mean
within column $n$ and $\sigma_{mn}$ the local variance within column
$n$, calculated with window sizes of 21 rowblocks. Note that
$\sigma_{mn}$ is calculated after the mean has been subtracted from
every rowblock background value along the column.

The bottom right panel of \fig{fig:obj} shows the resulting values of
$d_{mn}$ after application of this filter to the image shown in the
bottom left panel. We then apply an algorithm
\citep{1980CompJ..23..262L}\footnote{This is the same algorithm used
  in SExtractor \citep{1996A&AS..117..393B} to detect objects.} to
find \emph{objects} in the normalized background image. We define an
object as either a set of 20~or more adjoining pixels with values of
$d_{mn}$ larger than 1.5, or as a set of 5~or more pixels with values
larger than 2.5.  We then flag any photometric measurements where a
column affected by a moving object lies within the aperture used in
the photometry. The removal of all flagged data points results in the
elimination of all false positive events caused by moving objects.

\subsection{Event Search}

\subsubsection{Rank Product Statistic}
\label{sec:rank}
Our event selection algorithm uses the \emph{rank product statistic},
which is described in detail in \citet{2010PASP..122..959L}. This
statistic takes advantage of the multi-telescope data to give an exact
value for the significance of any candidate event, even if the
underlying distribution of the flux measurements is not known. Given
that TAOS samples at a 5~Hz readout cadence, we cannot resolve the
features of any candidate events in our lightcurves. Any occultation
event we detect would typically appear as a small drop in the flux for
one or two consecutive measurements (although occultation events by
larger objects could contain up to five or six consecutively measured
flux drops). We therefore need to have an accurate estimate of our
false positive rate in order to report a credible result.

\begin{figure*}
\plotone{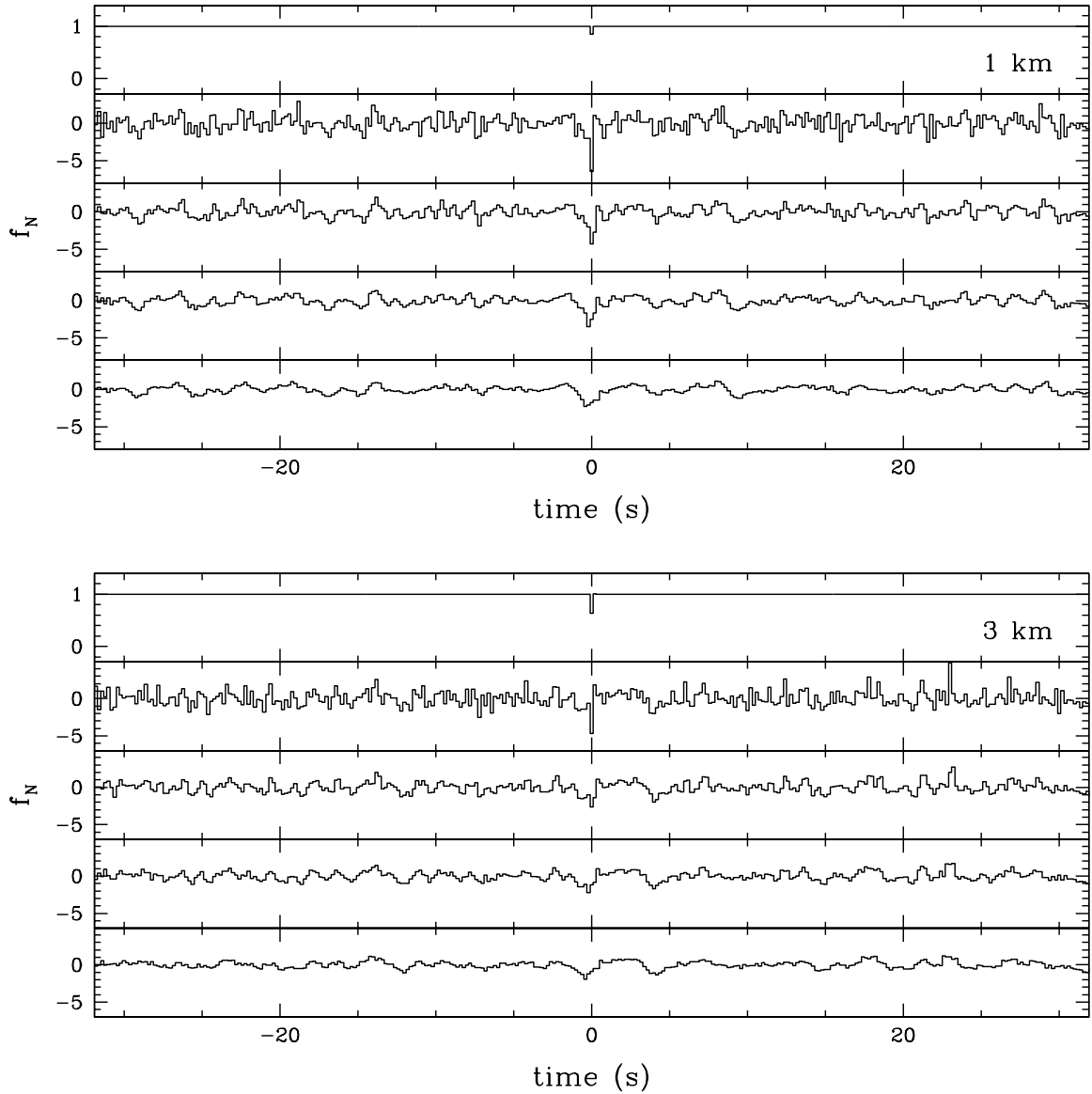}
\caption[]{Lightcurves with simulated occultation events from 1~km
  (top) and 3~km (bottom) diameter objects added in, before and after
  application of the moving average. In each panel, the top lightcurve
  is the simulated event, the second lightcurve has no moving average
  applied, and the next three have moving averages with window sizes
  2, 3, and 5 respectively.}
\label{fig:lc1}
\end{figure*}

\begin{figure*}
\plotone{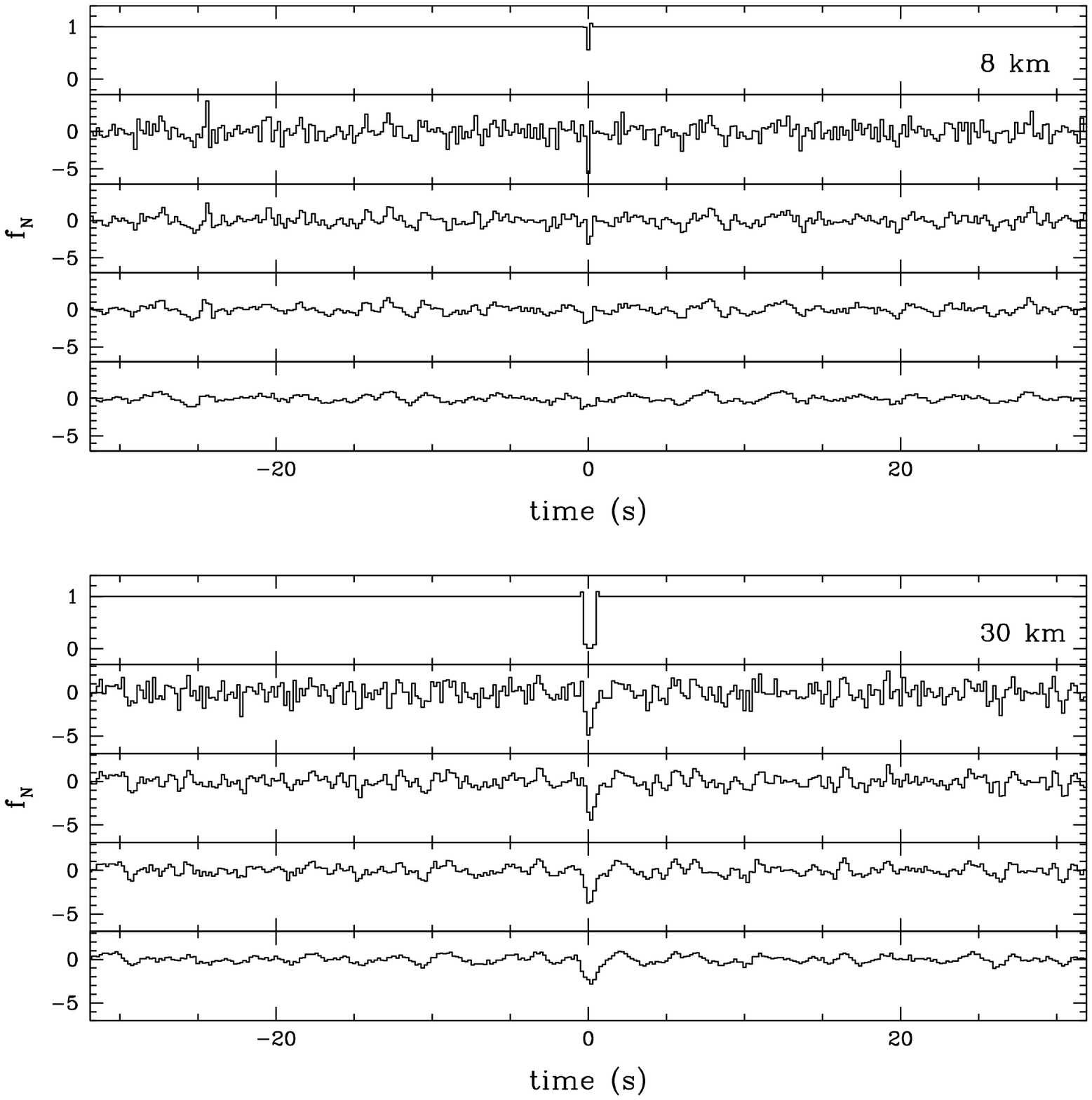}
\caption[]{Same as \fig{fig:lc1}, but for 8~km (top) and 30~km
  (bottom) diameter objects.}
\label{fig:lc2}
\end{figure*}

To search for events in the data, we calculate the rank product
statistic on each tuple in the entire data set. To calculate this
statistic, we take a time series of fluxes $f_1, f_2, \ldots, f_{\np}$
(where $\np$ is the number of points on the time series), and replace
each measurement $f_j$ with its rank $r_j$. The lowest measurement in
the time series will be given a rank of 1, and the brightest
measurement will be given a rank of $\np$. For a given lightcurve set,
we thus replace the time series of flux tuples with a time series of
\emph{rank tuples} of the form
\begin{displaymath}
(r_{1j}, \ldots, r_{Tj}),
\end{displaymath}
where $T$ is the number of telescopes used. We do not know the
underlying distribution of the flux tuples in each lightcurve set,
but, assuming the lightcurves meet the correct requirements (discussed
in \sect{sec:filt}), we do know the exact distribution of rank tuples,
and we can thus calculate the significance of any candidate event
exactly.

The rank product statistic is calculated as
\begin{equation}
z_j = -\ln \prod_{i=1}^T \frac{r_{ij}}{\np}.
\end{equation}
If one assumes that the rank distribution for a single lightcurve is
continuous, then this statistic follows the gamma distribution. Given
that the ranks are discrete, the approximation fails for large values
of $z$, however, the probability distribution can be calculated
exactly using simple combinatorics.

In the case of an occultation event, one expects the flux to drop
below the nominal value simultaneously in each of the telescopes. The
corresponding rank tuple would have lower values for all of the ranks,
and the resulting rank product would be significantly smaller than
average, leading to a large value of our test statistic $z$. We set a
threshold $\zt$ for event detection based on choosing an upper limit
on the expected false positive rate, such that
\begin{equation}
P(Z > \zt) = \frac{0.25}{\nt},
\label{eq:zt}
\end{equation}
where $Z$ is a random variable chosen from the rank product
probability distribution, 0.25 is our limit on the expected number
of false positive events in the data set, and $\nt$ is the
total number of tuples used in the event search.

\subsubsection{Lightcurve Filtering and Data Quality Checks}
\label{sec:filt}
The rank product test statistic follows the expected distribution for
an individual lightcurve set if and only if the following conditions
are satisfied \citep{nate}.
\begin{itemize}
\item The distribution of measurements in each lightcurve in the
  lightcurve set is stationary.
\item Each lightcurve in the lightcurve set is ergodic in mean.
\item Each lightcurve in the lightcurve set is independent of the
  other lightcurves in the lightcurve set.
\end{itemize}
However, none of the lightcurve sets in our data set meet these
requirements. Changes in the transparency of the atmosphere throughout
the course of our typical 1.5~hour data runs induce fluctuations in
the lightcurves, rendering them non-stationary. Furthermore, these
fluctuations are correlated between the different telescopes, so the
lightcurves are not independent of each other. We therefore apply a
rolling mean and variance filter to each point in the lightcurves,
thus replacing each flux measurement $f_j$ with $h_j$, given by
\begin{eqnarray}
g_j & = & f_j - \mu_j,\nonumber\\
h_j & = & \frac{g_j}{\sigma_j}.
\end{eqnarray}
where $\mu_j$ is the local mean and $\sigma_j$ the local variance at
point $j$, calculated with window sizes of 33 and 151
respectively. Note that the variance is calculated after the rolling
mean was subtracted, as was done in \eqn{eq:roll}. In most cases, this
filter removes all of the slowly varying trends, and the resulting
filtered lightcurves $h$ meet the requirements for the application of
the rank product statistic.

\begin{figure}
\plotone{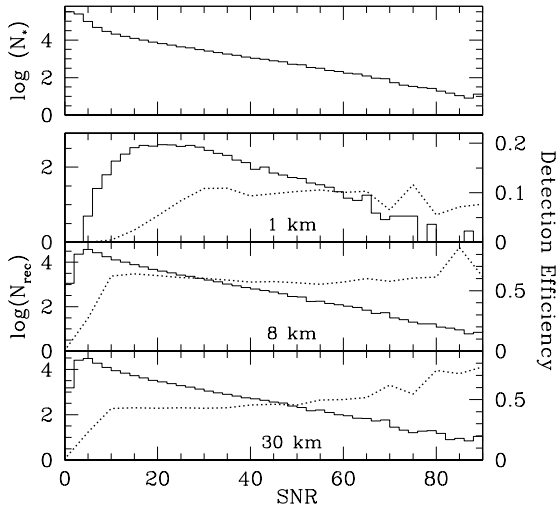}
\caption[]{Top panel: histogram of SNR values for lightcurve sets in
  the entire data set. (For each lightcurve set, the minimum SNR among
  all telescopes is used for the histogram.) Lower panels: SNR values
  of recovered events from our efficiency simulation for 1~km, 8~km,
  and 30~km objects (solid histogram). Also shown, using the right
  vertical axis, is the detection efficiency vs SNR (dotted lines).}
\label{fig:snr}
\end{figure}

However, there are still some data runs where fast moving cirrus
clouds can induce simultaneous variations in the lightcurves that are
not removed by the high pass filter. We have therefore devised a pair
of tests \citep{2010PASP..122..959L} to identify lightcurve sets where
the filtered lightcurves are not independent (and likely not
stationary as well, given that the fluctuations in the lightcurves
from the cirrus clouds are not constant over time). The tests are
based on the fact that the rank tuples should be distributed uniformly
over the $T$-dimensional rank space (recall that $T$ is the number of
telescopes used). For the first test, we divide the $T$-dimensional
rank space uniformly into a $T$-dimensional grid. If the lightcurves
are independent, we expect that each cell in the grid would contain
the same number of rank tuples. We thus calculate the Pearson's
$\chi^2$ statistic on the number of tuples in each cell of the
grid. For the second test, we only look at the number of rank tuples
in the lowest ranked cell in each dimension. When looking at all of
the lightcurve sets in a given data run, the results from the first
test should follow the $\chi^2$ distribution, and the results from the
second test should follow the hypergeometric distribution. (However,
this is only true if the lightcurves are independent and identically
distributed, which is a stricter requirement than stationary. We have
thus devised a \emph{blockwise bootstrap} method to calculate the
expected distributions. We still call the tests the Pearson's $\chi^2$
test and the hypergeometric test.) We can thus calculate a p-value for
both tests for each data run to see how well the two test statistics
match the expected distributions. Histograms of these p-values are
shown in \fig{fig:stat} for each test. If the statistics match the
distribution, we would expect a uniform distribution of p-values in
the range of $[0,1]$. The large spikes evident in the lowest bins of
each histogram are the data runs that have correlated fluctuations in
the lightcurves that are not removed by the high pass filter. This
leads to a non-uniform distribution of rank tuples, meaning that the
test statistics for each of the lightcurve sets does not follow the
expected distribution. We thus remove every data set with a p-value
less than 0.01. Note that this means we would only expect to remove
1\%~of our data runs due to random chance.

In order to perform these tests, we require enough high (SNR stars to
accurately calculate these two test statistics. We thus set a minimum
of 10~stars with $\mathrm{SNR} > 10$ in the data run. Any data run
that does not meet this requirement is excluded from further
analysis. Visual inspection of lightcurves in such data runs indicate
that there are very few stars, and the SNR is compromised, usually by
very bright sky during a full moon. These data runs are not expected
to contribute significantly to our effective sky coverage, so we lose
little by excluding them from the final results.  Furthermore, as will
be discussed in \sect{sec:snr}, most of the stars in these data runs
will end up being excluded anyway.

\begin{figure}
\plotone{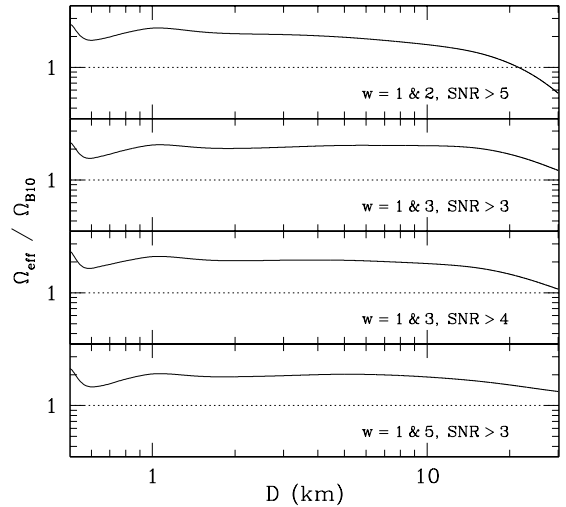}
\caption[]{Plots of $\aeff$ vs $D$ for the optimum combinations of
  moving average window sizes and SNR cuts. The values are scaled by
  $\Omega_\mathrm{B10}$, which is $\aeff$ from
  \citet{2010AJ....139.1499B}, in order to highlight the differences.}
\label{fig:oc}
\end{figure}

\subsubsection{Detection of Multipoint Occultation Events}
The rank product statistic described in \sect{sec:rank} only works on
one rank tuple at a time. This has the disadvantage of being
inefficient at detection of longer duration occultation events (such
as objects with $D \gtrsim 5$~km, scattered disk or Oort Cloud objects
beyond 100~AU, or events measured at large opposition
angles). However, as was shown in \citet{nate} and discussed in
\citet{2010PASP..122..959L}, if we replace a stationary lightcurve
with a moving average of the form
\begin{equation}
a_j = \frac{1}{w}(h_{j} + \ldots + h_{j+w-1}),
\end{equation}
where $h_j$ is the lightcurve after high pass filtering and $w$ is the
rolling average window size, then the resulting lightcurve will also
be stationary, and the rank product statistic will still be
applicable. By applying a moving average, we can increase the
significance of any candidate event since the average lightcurve will
exhibit a more significant drop in the measured flux when the averaging
window is centered on the event.

This is illustrated in Figures~\ref{fig:lc1} and~\ref{fig:lc2}, which
shows the results of the application of a moving average on actual
lightcurves with simulated events added in. We have used three
different window sizes $w$ = 2, 3, and 5, in order to optimize our
detection efficiency. We also looked at the lightcurves with no
average applied, which we will refer to as using a window size $w = 1$
in order to simplify the discussion. For the 1~km diameter simulated
event, we detect the event using $w = 1$, 2, and 3. For such a small
object, using $w = 5$ actually smoothes out the signal, while
amplifying other small fluctuations, so that the event does not pass
the cuts. For the 3~km and 8~km diameter objects, the events are
detected only with $w = 1$ and~2. For the 30~km object, which is
longer in duration than the other simulated events, the signal is not
detectable for $w = 1$, but it is detected with $w = 2$, 3, and 5.

\begin{figure}
\plotone{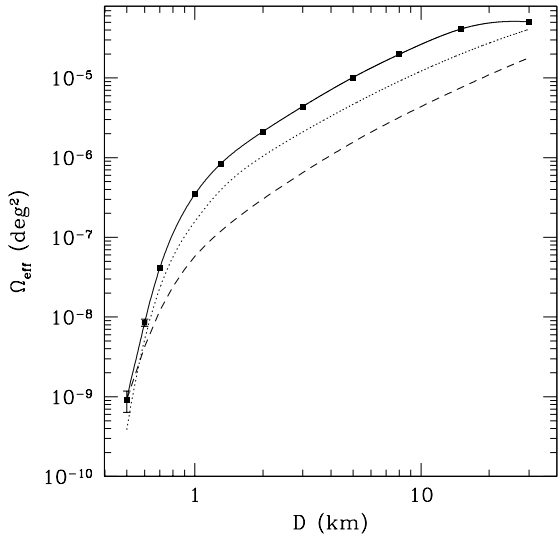}
\caption[]{$\aeff$ vs $D$ using our final selection criteria of $w = 1$
  and 3 and $\mathrm{SNR} > 3$. Also shown for comparison are the
  previous TAOS results from \citet{2010AJ....139.1499B} (dotted line)
  and \citet{2008ApJ...685L.157Z} (dashed line).}
\label{fig:omega}
\end{figure}

It is clear that the different window sizes increase the sensitivity
to different size objects. However, while the effective sky coverage
of the survey can be increased by using a wide array of window sizes,
this will also increase the false positive rate
\citep{2010PASP..122..959L}. Running each multiple window size on each
lightcurve set is not an independent test, given that events can be
detected using more than one window size, as demonstrated in
Figures~\ref{fig:lc1} and~\ref{fig:lc2}. It follows that any false
positive event could be detected multiple times using different window
sizes. However, given that we do not know the underlying distribution
of flux values in the data set, we can not estimate the rate at which
this may occur. Therefore, instead of estimating the true false
positive rate, we can only set an upper limit on the false positive
rate by assuming that searching for events in a lightcurve set after
applying the moving average is a completely independent test for each
value of $w$. If we have a total of $N_\mathrm{w}$ window sizes, we
thus modify \eqn{eq:zt} as
\begin{equation}
P(Z > \zt) = \frac{0.25}{\nt N_\mathrm{w}}.
\label{eq:ztw}
\end{equation}
So the more window sizes that are used, the tighter the detection
threshold must be. The optimization of the selection of the window
sizes used in this analysis will be discussed in \sect{sec:results}.

Note that the data quality tests described in \sect{sec:filt} must be
reapplied for each window size $w$. As shown in
\citet{2010PASP..122..959L}, application of the moving average can
highlight small correlations between the telescopes that are
insignificant in the unaveraged data. Therefore using larger window
sizes will require removing more data runs from the analysis. In cases
where we use more than one window size, we always choose to be
conservative and remove the union of data sets that would be rejected
using each value of $w$.

\subsubsection{Signal to Noise Ratio Cuts}
\label{sec:snr}
The final cut we apply to our data set is to set a threshold on SNR of
each lightcurve set. The goal of this cut is to exclude stars which
provide little sensitivity to occultation events, yet still contribute
to the total number of tuples $\nt$, which tightens the selection
threshold (see \eqn{eq:ztw}). This is illustrated in \fig{fig:snr},
which shows the total distribution of SNR values for every lightcurve
set in the data set, as well as the SNR distributions of stars for
which simulated events are recovered for 1~km, 8~km, and 30~km
objects. It is clear that there is little contribution to our
effective sky coverage for $\mathrm{SNR} \lesssim 5$. Also shown is
the detection efficiency for the same diameter objects vs. SNR. As
expected, the detection efficiency drops off at low SNR values,
especially for the smaller objects.

The exact value of SNR to use as a threshold depends on many factors,
especially the set of moving average window sizes that will be
used. This optimization is discussed in \sect{sec:results}.

\begin{deluxetable*}{ccccccc}
\tablecolumns{7}
\tablewidth{0pc}
\tablecaption{Results from using different window sizes and SNR cuts.}
\tablehead{Window Sizes & SNR Cut & Tuples & $z_\mathrm{t}$ & $q$ & 
$\aeff(D=1\,\mathrm{km})$ & $\aeff(D=30\,\mathrm{km})$}
\startdata
1 \& 2 & 3 & $9.49\times10^{9}$ & $1.32\times 10^{-11}$ & 3.82 &
$(3.590\pm0.053)\times10^{-7}$ & $(2.788\pm0.011)\times10^{-5}$ \\
1 \& 2 & 4 & $7.13\times10^{9}$ & $1.75\times 10^{-11}$ & 3.81 &
$(3.703\pm0.054)\times10^{-7}$ & $(2.577\pm0.010)\times10^{-5}$ \\
1 \& 2 & 5 & $5.63\times10^{9}$ & $2.22\times 10^{-11}$ & 3.82 &
$(3.806\pm0.055)\times10^{-7}$ & $(2.277\pm0.010)\times10^{-5}$ \\
1 \& 3 & 3 & $9.14\times10^{9}$ & $1.37\times 10^{-11}$ & 3.82 &
$(3.467\pm0.053)\times10^{-7}$ & $(5.065\pm0.014)\times10^{-5}$ \\
1 \& 3 & 4 & $6.87\times10^{9}$ & $1.82\times 10^{-11}$ & 3.82 &
$(3.559\pm0.053)\times10^{-7}$ & $(4.430\pm0.013)\times10^{-5}$ \\
1 \& 3 & 5 & $5.42\times10^{9}$ & $2.31\times 10^{-11}$ & 3.82 &
$(3.657\pm0.054)\times10^{-7}$ & $(3.770\pm0.012)\times10^{-5}$ \\
1 \& 5 & 3 & $8.60\times10^{9}$ & $1.45\times 10^{-11}$ & 3.84 &
$(3.213\pm0.051)\times10^{-7}$ & $(5.594\pm0.015)\times10^{-5}$ \\
1 \& 5 & 4 & $6.46\times10^{9}$ & $1.94\times 10^{-11}$ & 3.84 &
$(3.301\pm0.052)\times10^{-7}$ & $(4.824\pm0.014)\times10^{-5}$ \\
1 \& 5 & 5 & $5.08\times10^{9}$ & $2.46\times 10^{-11}$ & 3.84 &
$(3.413\pm0.052)\times10^{-7}$ & $(4.071\pm0.013)\times10^{-5}$ \\
\enddata
\label{tbl:results}
\end{deluxetable*}

\begin{figure*}
\plotone{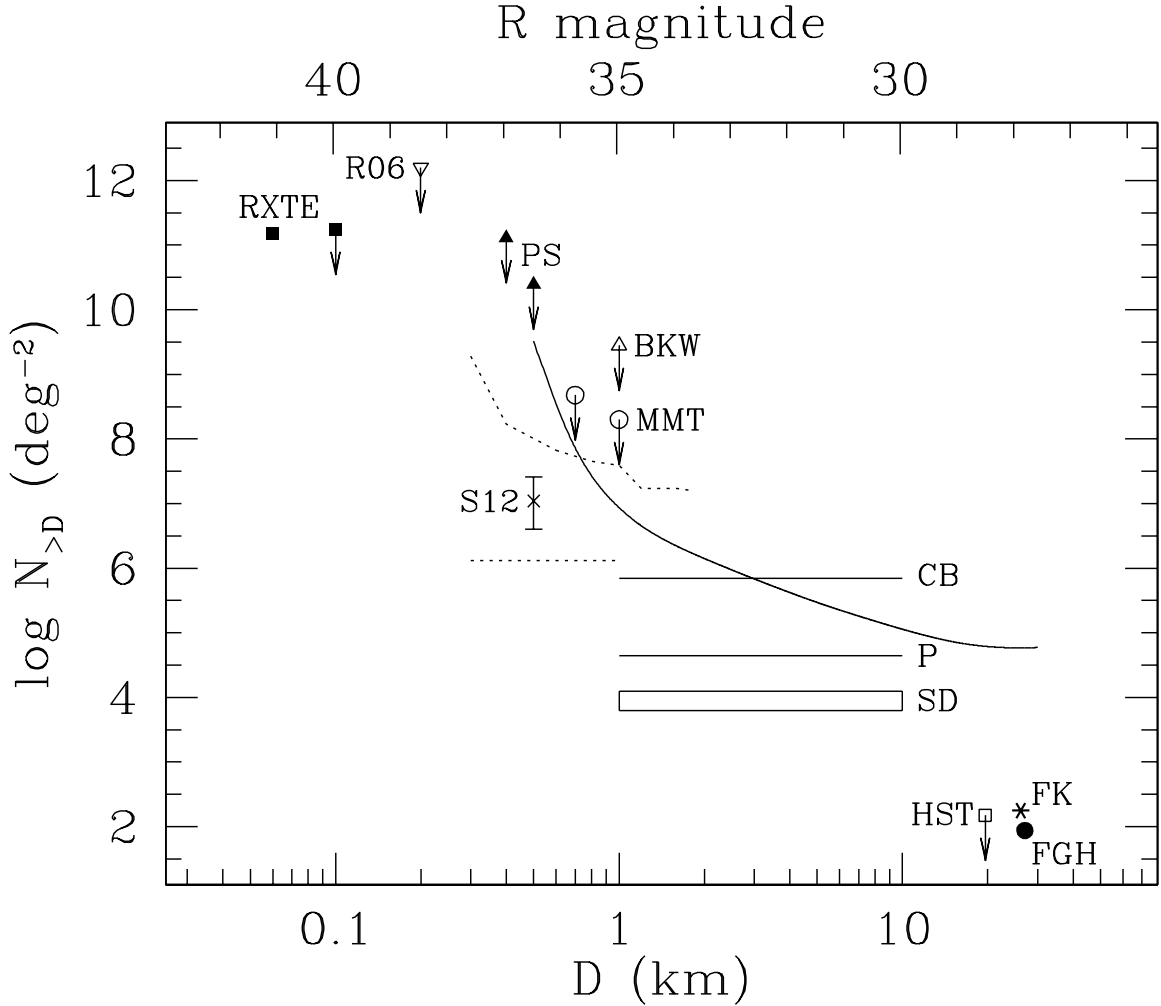}
\caption[]{Solid line: 95\% c.l. upper limit on cumulative surface
  density vs. diameter $D$ (bottom axis) and $R$~magnitude (top axis,
  assuming an albedo of 4\% and a distance of 43~AU) from current TAOS
  data set. Dotted lines: 95\%~c.l. upper and lower limits on surface
  density from \citet{2012ApJ...761..150S}. Cross and error bar (S12):
  Surface density reported by \citet{2012ApJ...761..150S}. (Note that
  this is not strictly model independent because it is based on a
  power law model. However, the result does not depend very strongly
  on the assumed slope.) Solid squares (RXTE): upper limit (right point)
  reported by \citet{2008MNRAS.388L..44L} and the best fit surface
  density (left point) reported by \citet{2007MNRAS.378.1287C} from an
  occultation search through RXTE observation of Sco~X-1. Solid
  triangles (PS): upper limits reported by \citet{2010AJ....139.2003W}
  using Pan-STARRS guider images. Empty triangle (BKW): upper limit
  reported by \citet{2008AJ....135.1039B} using the 1.8~m telescope at
  DAO. Empty circles (MMT): upper limits reported by
  \citet{2009AJ....138..568B} from analysis of trailed images obtained
  with the MMT. Upside down empty triangle (R06): upper limit reported
  by \cite{2006AJ....132..819R} from high speed imaging using the
  4.2~m Herschel Telescope. Empty square (HST): upper limit reported
  by \citet{2004AJ....128.1364B} from a direct survey using the
  HST. Solid circle (FGH): surface density reported by
  \citet{2009ApJ...696...91F} from a direct search using Subaru. Star
  (FK): surface density reported by \citet{2009AJ....137...72F} from a
  direct survey using Subaru. (Note that \citet{2009AJ....137...72F}
  assume an albedo of 6\% and a distance of 35~AU, so they claim that
  the point plotted at $R = 27.875$ corresponds to a diameter $D =
  15$~km.) Lines and box labeled CB, P, and SD: required surface
  density for the Classical Belt \citep{1997Icar..127...13L}, Plutinos
  \citep{1997Icar..127....1M}, and Scattered Disk
  \citep{2008ApJ...687..714V} respectively to be the source for the
  observed distribution of Jupiter family comets.}
\label{fig:lim}
\end{figure*}

\section{Results}
\label{sec:results}
We searched through our data set, using moving average window sizes of
1, 2, 3, and 5, (and every combination thereof) as well as SNR cuts of
0, 1, 2, 3, 4, 5, 6, and 7. For each combination of window size and SNR
threshold, the detection threshold $\zt$ was calculated as
shown in \eqn{eq:ztw}. In all cases, no events were found in the data
set. So the next step in the analysis was to measure our effective sky
coverage and optimize the selection of moving average window sizes and
SNR cut.

The total number of events expected by the survey if given by
\begin{equation}
N_\mathrm{exp} = \int\frac{dn}{dD}\,\aeff(D)\,dD,
\label{eq:nexp}
\end{equation}
where $dn/dD$ is the differential size distribution (the number of
objects per deg$^{-2}$ per km along the ecliptic), and $\aeff$ is the
effective sky coverage of the survey. We estimate $\aeff$ by
implanting a simulated event into each lightcurve set and seeing if it
is recovered by our selection criteria. For this simulation, we choose
a set of diameters between 0.5~km and 30~km, and for each diameter, we
implant exactly one event into each lightcurve set. We assume every
object is at a geocentric distance of 43~AU, since the lightcurve
shape does not depend critically on the distance for most objects
(40~to 46~AU) in the Kuiper Belt. The tests for each diameter are
performed independently, so there is never more than one event
implanted into any lightcurve set at any time. The effective sky
coverage is calculated as
\begin{equation}
\aeff(D) = \sum_{l \in \mathrm{rec}}
\frac{H}{\Delta}~\frac{v_\mathrm{rel}}{\Delta}~E_l,
\end{equation}
where the sum is over only those lightcurve sets $l$ where the added
event was recovered, $E_l$ is the length of the lightcurve set in
time, $v_\mathrm{rel}$ is the relative transverse velocity between the
Earth and the KBO, $\Delta$ is the geocentric distance to the KBO
(again, fixed to a constant value of 43~AU), and $H$ is the event
cross section. We estimate $H$ as
\begin{equation}
H(D,\Delta,\theta_*) \approx \left[(2\sqrt{3}F^{\frac{3}{2}}
  + D^{\frac{3}{2}} \right]^{\frac{2}{3}} + \theta_*\Delta,
\label{eq:h}
\end{equation}
where $\theta_*$ is the angular size of the target star and $F$ is the
Fresnel scale, which is included to account for the minimum cross
section of an event due to diffraction \citep{2007AJ....134.1596N}.

To optimize the selection of window sizes and SNR cut, we looked at
the resulting values of $\aeff$ for every possible combination. There
is no clear optimum set of cuts, so we decided to optimize our
selection based on three criteria.

\begin{itemize}
\item Minimize the upper limit of the slope $q$ at the small end of
  the size distribution, assuming a power law size distribution
  anchored at the break diameter of 90~km and a cumulative surface
  density at the break diameter of 5.4~deg$^{-2}$
  \citep{2009AJ....137...72F, 2009Natur.462..895S}.
\item Maximize the sensitivity at $D = 1$~km, where
  \citet{2009Natur.462..895S} and \citet{2012ApJ...761..150S} claim
  the detection of two KBOs.
\item Maximize the sensitivity at $D = 30$~km, in order to bring our
  upper limit as close as possible to the current direct detection
  limits \citep{2004AJ....128.1364B, 2009ApJ...696...91F,
    2009AJ....137...72F}.
\end{itemize}

The best results came from using cuts on SNR of 3, 4, and 5. The
results are summarized in \tbl{tbl:results}. We found that the upper
limit on the slope $q$ did not vary significantly as long as we
limited ourselves to two window sizes, and included $w = 1$. The
combination of $w = 1$ and 2 and a minimum of $\mathrm{SNR} > 3$ gave
the largest effective sky coverage at 1~km, however, this was
particularly bad at 30~km. The best result at 30~km came from $w = 1$
and 5 with $\mathrm{SNR} > 3$, but this combination was not very good
for 1~km. The combinations that worked best for both diameters was
using $w = 1$ and 3 with SNR cuts of 3 and 4. The resulting values of
$\aeff$ for these four combinations are shown in \fig{fig:oc}. In the
end, we opted for using $w = 1$ and 3, with a threshold of
$\mathrm{SNR} > 3$ as this is better at 30~km, and not much worse at
1~km.
 
Given our final set of cuts, the resulting plot of $\aeff$ vs $D$ is
shown \fig{fig:omega}. The points on the curve indicate the diameters
where we calculated $\aeff$ using our detection efficiency simulation,
and the curve itself is a cubic spline fit to these values.

To set a model independent upper limit, with no detected events we can
eliminate any model which would lead us to expect more that three
detected events at the 95\% c.l. Our model independent upper limit is
shown in \fig{fig:lim}, along with results from other occultation
surveys and direct searches. We note that our upper limit at $D =
1$~km is a factor of 4.5 below that reported by
\citet{2012ApJ...761..150S}, but it is also a factor of 6.7 higher
than their lower limit based upon their two reported events. Therefore
even though we found no events at 1~km (despite the fact that our
efficiency simulation for 1~km diameter objects showed that we are in
fact sensitive to events similar to those detected by the HST survey),
our limit is thus still consistent with their published result.

\begin{figure}
\plotone{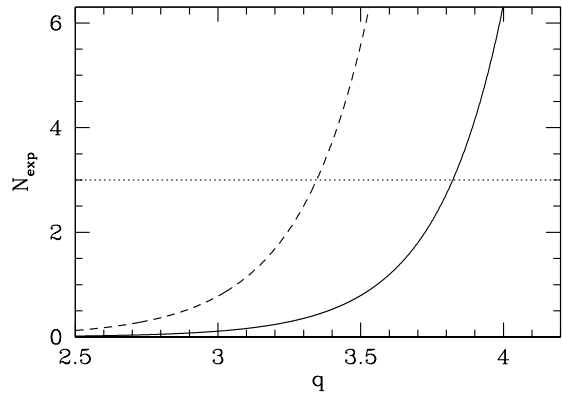}
\caption[]{Number of expected events vs slope $q$. Solid line: number
  of expected events using a surface density of $N_{D>90\;\mathrm{km}}
  = 5.4$~deg$^{-2}$, as reported by
  \citet{2009AJ....137...72F}. Dashed line: number of expected events
  vs. slope $q$, using a surface density of $N_{D>90\;\mathrm{km}} =
  38.0$~deg$^{-2}$, as reported by \citet{2009ApJ...696...91F}. Since
  no events were found, any model predicting more than 3~events
  (dotted line) is excluded at the 95\% c.l.}
\label{fig:nvsq}
\end{figure}

We also calculate upper limits on the small KBO size distribution
assuming it follows a power law anchored at the detected break at $D =
90$~km. We use \eqn{eq:nexp} to calculate the number of expected
events as a function of slope $q$. The results are shown in
\fig{fig:nvsq}. We use two values for the surface density at $D =
90$~km, which normalize the size distribution for smaller
diameters. Using the value reported by \citet{2009AJ....137...72F} of
$N_{D>90\;\mathrm{km}} = 5.4$~deg$^{-2}$ yields a 95\%~c.l. upper
limit of $q = 3.82$. In comparison, \citet{2009Natur.462..895S}
report a slope of $q = 3.8 \pm 0.2$ based on their claim of two
detections at $D = 1$~km, assuming the inclination distribution of
small KBOs follows that of the larger ($D > 100$~km) objects
\citep{2005AJ....129.1117E}. The second value we use is
$N_{D>90\;\mathrm{km}} = 38.0$~deg$^{-2}$ as reported by
\citet{2009ApJ...696...91F}. This results in a 95\%~c.l. upper limit
on $q$ of 3.34.

\section{Future Plans}
\label{sec:plans}
After seven years of observations, it would only be of marginal value
to continue the survey in its present form. So at the end of the data
set discussed in this paper, we shut down the system for a camera
upgrade. Each new camera, manufactured by Spectral Instruments,
utilizes a CCD47-20 frame transfer CCD from e2v. With the new cameras,
we can image with a readout cadence of just under 10~Hz if we use
2$\times$2 binning. The CCDs are 1k$\times$1k with 13~$\mu$m pixels,
as opposed to the 2k$\times$2k imagers with 13.5~$\mu$m pixels used to
collect the current data set, so 77\% of the field of view is
lost. However, by moving away from zipper mode to full frame imaging,
our SNR increases significantly and we estimate that our limiting
magnitude will increase from $R = 13.5$ to 15. We thus expect to be
able to monitor a similar number of stars. Given the higher readout
cadence, we become significantly more sensitive to objects with $D
\lesssim 1$~km. Observations with these new cameras began in November
2012.

\acknowledgements Work at ASIAA was supported in part by the thematic
research program AS-88-TP-A02. Work at NCU and at Lulin Observatory was
supported in part by grant NSC101-2628-M-008-002. Work at the CfA
was supported in part by the NSF under grant AST-0501681 and by NASA
under grant NNG04G113G. Work at Yonsei was supported by the NRF grant
2011-0030875. Work at NCU was supported by the grant NSC
96-2112-M-008-024-MY3. Work at Berkeley was supported in part by NSF
grant DMS-0636667. Work at SLAC was performed under USDOE contract
DE-AC02-76SF00515. Work at NASA Ames was supported by NASA's Planetary
Geology \& Geophysics Program.

\bibliography{ms}

\begin{thebibliography}{55}
\expandafter\ifx\csname natexlab\endcsname\relax\def\natexlab#1{#1}\fi

\bibitem[{{Bailey}(1976)}]{1976Natur.259..290B}
{Bailey}, M.~E. 1976, \nat, 259, 290

\bibitem[{{Benavidez} \& {Campo Bagatin}(2009)}]{2009P&SS...57..201B}
{Benavidez}, P.~G., \& {Campo Bagatin}, A. 2009, Planet. Space Sci., 57, 201

\bibitem[{{Benz} \& {Asphaug}(1999)}]{1999Icar..142....5B}
{Benz}, W., \& {Asphaug}, E. 1999, Icarus, 142, 5

\bibitem[{{Bernstein} {et~al.}(2004)}]{2004AJ....128.1364B}
{Bernstein}, G.~M., {et~al.} 2004, AJ, 128, 1364

\bibitem[{{Bertin} \& {Arnouts}(1996)}]{1996A&AS..117..393B}
{Bertin}, E., \& {Arnouts}, S. 1996, A\&AS, 117, 393

\bibitem[{{Bianco} {et~al.}(2009)}]{2009AJ....138..568B}
{Bianco}, F.~B., {et~al.} 2009, AJ, 138, 568

\bibitem[{{Bianco} {et~al.}(2010)}]{2010AJ....139.1499B}
---. 2010, AJ, 139, 1499

\bibitem[{{Bickerton} {et~al.}(2008){Bickerton}, {Kavelaars}, \&
  {Welch}}]{2008AJ....135.1039B}
{Bickerton}, S.~J., {Kavelaars}, J.~J., \& {Welch}, D.~L. 2008, AJ, 135, 1039

\bibitem[{{Bickerton} {et~al.}(2009){Bickerton}, {Welch}, \&
  {Kavelaars}}]{2009AJ....137.4270B}
{Bickerton}, S.~J., {Welch}, D.~L., \& {Kavelaars}, J.~J. 2009, \aj, 137, 4270

\bibitem[{{Blocker} {et~al.}(2009){Blocker}, {Protopapas}, \&
  {Alcock}}]{2009ApJ...701.1742B}
{Blocker}, A.~W., {Protopapas}, P., \& {Alcock}, C.~R. 2009, \apj, 701, 1742

\bibitem[{{Brown} \& {Webster}(1997)}]{1997MNRAS.289..783B}
{Brown}, M.~J.~I., \& {Webster}, R.~L. 1997, \mnras, 289, 783

\bibitem[{{Chang} {et~al.}(2012){Chang}, {Liu}, \&
  {Chen}}]{2012MNRAS.tmp..367C}
{Chang}, H.-K., {Liu}, C.-Y., \& {Chen}, K.-T. 2012, \mnras, 367

\bibitem[{{Chang} {et~al.}(2006)}]{2006Natur.442..660C}
{Chang}, H.-K., {et~al.} 2006, \nat, 442, 660

\bibitem[{{Chang} {et~al.}(2007)}]{2007MNRAS.378.1287C}
---. 2007, \mnras, 378, 1287

\bibitem[{{Coehlo}(2010)}]{nate}
{Coehlo}, N.~K. 2010, PhD thesis, University of California, Berkeley

\bibitem[{{Cooray}(2003)}]{2003ApJ...589L..97C}
{Cooray}, A. 2003, ApJL, 589, L97

\bibitem[{{Cooray} \& {Farmer}(2003)}]{2003ApJ...587L.125C}
{Cooray}, A., \& {Farmer}, A.~J. 2003, ApJL, 587, L125

\bibitem[{{Davis} \& {Farinella}(1997)}]{1997Icar..125...50D}
{Davis}, D.~R., \& {Farinella}, P. 1997, Icarus, 125, 50

\bibitem[{{Duncan} \& {Levison}(1997)}]{1997Sci...276.1670D}
{Duncan}, M.~J., \& {Levison}, H.~F. 1997, Science, 276, 1670

\bibitem[{{Duncan} {et~al.}(1995){Duncan}, {Levison}, \&
  {Budd}}]{1995AJ....110.3073D}
{Duncan}, M.~J., {Levison}, H.~F., \& {Budd}, S.~M. 1995, AJ, 110, 3073

\bibitem[{{Elliot} {et~al.}(2005)}]{2005AJ....129.1117E}
{Elliot}, J.~L., {et~al.} 2005, \aj, 129, 1117

\bibitem[{{Fraser} \& {Kavelaars}(2008)}]{2008Icar..198..452F}
{Fraser}, W.~C., \& {Kavelaars}, J.~J. 2008, Icarus, 198, 452

\bibitem[{{Fraser} \& {Kavelaars}(2009)}]{2009AJ....137...72F}
---. 2009, AJ, 137, 72

\bibitem[{{Fraser} {et~al.}(2008)}]{2008Icar..195..827F}
{Fraser}, W.~C., {et~al.} 2008, Icarus, 195, 827

\bibitem[{{Fuentes} {et~al.}(2009){Fuentes}, {George}, \&
  {Holman}}]{2009ApJ...696...91F}
{Fuentes}, C.~I., {George}, M.~R., \& {Holman}, M.~J. 2009, ApJ, 696, 91

\bibitem[{{Fuentes} \& {Holman}(2008)}]{2008AJ....136...83F}
{Fuentes}, C.~I., \& {Holman}, M.~J. 2008, AJ, 136, 83

\bibitem[{{Holman} \& {Wisdom}(1993)}]{1993AJ....105.1987H}
{Holman}, M.~J., \& {Wisdom}, J. 1993, AJ, 105, 1987

\bibitem[{{Jones} {et~al.}(2008)}]{2008ApJ...677.1241J}
{Jones}, T.~A., {et~al.} 2008, ApJ, 677, 1241

\bibitem[{{Kenyon} \& {Bromley}(2001)}]{2001AJ....121..538K}
{Kenyon}, S.~J., \& {Bromley}, B.~C. 2001, AJ, 121, 538

\bibitem[{{Kenyon} \& {Bromley}(2004)}]{2004AJ....128.1916K}
---. 2004, AJ, 128, 1916

\bibitem[{{Kenyon} \& {Bromley}(2009)}]{2009ApJ...690L.140K}
---. 2009, ApJL, 690, L140

\bibitem[{{Kenyon} \& {Luu}(1999{\natexlab{a}})}]{1999AJ....118.1101K}
{Kenyon}, S.~J., \& {Luu}, J.~X. 1999{\natexlab{a}}, AJ, 118, 1101

\bibitem[{{Kenyon} \& {Luu}(1999{\natexlab{b}})}]{1999ApJ...526..465K}
---. 1999{\natexlab{b}}, ApJ, 526, 465

\bibitem[{{Lehner} {et~al.}(2009)}]{2009PASP..121..138L}
{Lehner}, M.~J., {et~al.} 2009, PASP, 121, 138

\bibitem[{{Lehner} {et~al.}(2010)}]{2010PASP..122..959L}
---. 2010, PASP, 122, 959

\bibitem[{{Levison} \& {Duncan}(1997)}]{1997Icar..127...13L}
{Levison}, H.~F., \& {Duncan}, M.~J. 1997, Icarus, 127, 13

\bibitem[{{Liu} {et~al.}(2008)}]{2008MNRAS.388L..44L}
{Liu}, C.-Y., {et~al.} 2008, \mnras, 388, L44

\bibitem[{{Lutz}(1980)}]{1980CompJ..23..262L}
{Lutz}, R.~K. 1980, The Computer Journal, 23, 262

\bibitem[{{Luu} \& {Jewitt}(2002)}]{2002ARA&A..40...63L}
{Luu}, J.~X., \& {Jewitt}, D.~C. 2002, ARA\&A, 40, 63

\bibitem[{{Morbidelli}(1997)}]{1997Icar..127....1M}
{Morbidelli}, A. 1997, Icarus, 127, 1

\bibitem[{{Nihei} {et~al.}(2007)}]{2007AJ....134.1596N}
{Nihei}, T.~C., {et~al.} 2007, AJ, 134, 1596

\bibitem[{{Pan} \& {Sari}(2005)}]{2005Icar..173..342P}
{Pan}, M., \& {Sari}, R. 2005, Icarus, 173, 342

\bibitem[{{Roques} \& {Moncuquet}(2000)}]{2000Icar..147..530R}
{Roques}, F., \& {Moncuquet}, M. 2000, Icarus, 147, 530

\bibitem[{{Roques} {et~al.}(1987){Roques}, {Moncuquet}, \&
  {Sicardy}}]{1987AJ.....93.1549R}
{Roques}, F., {Moncuquet}, M., \& {Sicardy}, B. 1987, AJ, 93, 1549

\bibitem[{{Roques} {et~al.}(2003)}]{2003ApJ...594L..63R}
{Roques}, F., {et~al.} 2003, ApJL, 594, L63

\bibitem[{{Roques} {et~al.}(2006)}]{2006AJ....132..819R}
---. 2006, AJ, 132, 819

\bibitem[{{Schlichting} {et~al.}(2009)}]{2009Natur.462..895S}
{Schlichting}, H.~E., {et~al.} 2009, \nat, 462, 895

\bibitem[{{Schlichting} {et~al.}(2012)}]{2012ApJ...761..150S}
---. 2012, \apj, 761, 150

\bibitem[{{Stern}(1996)}]{1996AJ....112.1203S}
{Stern}, S.~A. 1996, AJ, 112, 1203

\bibitem[{{Tancredi} {et~al.}(2006)}]{2006Icar..182..527T}
{Tancredi}, G., {et~al.} 2006, Icarus, 182, 527

\bibitem[{{Volk} \& {Malhotra}(2008)}]{2008ApJ...687..714V}
{Volk}, K., \& {Malhotra}, R. 2008, ApJ, 687, 714

\bibitem[{{Wang} {et~al.}(2009)}]{2009AJ....138.1893W}
{Wang}, J.-H., {et~al.} 2009, AJ, 138, 1893

\bibitem[{{Wang} {et~al.}(2010)}]{2010AJ....139.2003W}
---. 2010, AJ, 139, 2003

\bibitem[{{Zhang} {et~al.}(2008)}]{2008ApJ...685L.157Z}
{Zhang}, Z.-W., {et~al.} 2008, ApJL, 685, L157

\bibitem[{{Zhang} {et~al.}(2009)}]{2009PASP..121.1429Z}
---. 2009, PASP, 121, 1429

\end{thebibliography}

\end{document}